\documentclass[aps,twocolumn,superscriptaddress]{revtex4-1}
\usepackage{graphics} 
\usepackage{epsfig} 
\usepackage{amsmath}
\usepackage{color}
\newcommand{\uvec}[1]{\boldsymbol{\hat{\textbf{#1}}}}
\usepackage[english]{babel} 

\usepackage{xmpmulti}

\usepackage{babel}
\begin{document}
\title{Study on Kelvin Helmholtz shear flows subjected to differential rotation  
}
\author{Prince Kumar}
\email{prince.kumar@ipr.res.in}
\affiliation{Institute for Plasma Research, Bhat, Gandhinagar, India, 382428}
\author{Devendra Sharma}
\email{devendra@ipr.res.in}
\affiliation{Institute for Plasma Research, Bhat, Gandhinagar, India, 382428}
\affiliation{Homi Bhabha National Institute, Training School Complex, Anushaktinagar, Mumbai 400094, India}
\date{\today}

\begin{abstract}
A numerical simulation of Kelvin-Helmholtz Instability (KHI) in parallel shear flows subjected to external rotation is carried out using a pseudo-spectral technique. The Coriolis force, arising in a rotation frame under the beta plane approximation, tends to suppress the growth of KHI modes. The numerical results show a close qualitative agreement with the analytical results obtained for a step-wise shear flow profile. Experimental evidence demonstrates that particles in a rotating frame experience the Coriolis force, mathematically equivalent to the Lorentz force. Therefore, the Coriolis force affects fluid dynamics in a manner similar to the Lorentz force in magnetized shear flows. This paper exploits the analogy between the magnetic field and rotation to study effects equivalent to a magnetic field on KHI in a rotating frame. Similar to the magnetic field case, the Coriolis force suppresses KHI and tends to form compressed and elongated KH vortex structures. However, the magnetic field and Coriolis force act on different scales, with the latter suppressing long-wavelength mode perturbations. A higher number of vortices are observed in the presence of rotation compared to non-rotating cases.

\end{abstract}
\maketitle
\section*{Introduction} 
Fluids exhibiting rotational motion are prevalent in numerous technical domains and play a significant role in geophysical phenomena, particularly within the atmospheric and oceanic realms \cite{lambeck2005earth,cushman2011introduction,sidorenkov2005physics}. The Coriolis and centrifugal forces, acting on large-scale rotating flows, significantly affect their characteristic features, particularly influencing the instabilities that arise under different physical conditions    \cite{vallis1993generation,fruh1999experiments,drazin1961discontinuous,read2020baroclinic}. Instabilities within flows that occur on a small or large scale are crucial mechanisms responsible for transporting momentum and energy and influencing the evolution of many astrophysical objects \cite{miura1982nonlocal}. The Kelvin-Helmholtz Instability (KHI) arises in velocity shear flows and is a well-known type of instability observed in a wide range of natural systems, such as the Earth's atmosphere, ocean waves, and astrophysical jets \cite{chandrasekhar2013hydrodynamic,drazin2002introduction}. The growth rate of the instability is influenced by factors such as surface tension, magnetic field, rotation, density, and gravity \cite{chandrasekhar2013hydrodynamic,keppens1999growth}. The effects of a magnetic field on the characteristics of KHI have been extensively studied, revealing that a magnetic field aligned with the flow direction suppresses KHI instability \cite{liu2018physical,keppens1999growth}. However, a transverse magnetic field does not affect the development of KHI . The stabilizing effects
of the magnetic field is known to originate from the Lorentz force\cite{liu2018physical}.

Interestingly, rotation has been identified as a source that mimics the effects of a strong magnetic field, where the Coriolis force acts like a Lorentz force of magnetized systems \cite{kumar2021collective, hartmann2013,kahlert2012magnetizing}. In a recent experiment\cite{hartmann2013,kahlert2012magnetizing}, it has been demonstrated that the constituent particles experience the same deflection as they would in the presence of a magnetic field when measurements are carried out in a rotating frame. The purpose of the paper is to study the effects of the Coriolis force on the Kelvin-Helmholtz instability and attempt to identify the analogy with a magnetic field in large-scale flows. The vorticity in inviscid fluids is altered solely by the magnetic and Coriolis forces in the absence of pressure gradient forces. Due to these forces, vorticity is present in the oceans, atmosphere, and magneto-hydrodynamic flows. \\ 
The Coriolis force effects are introduced in the analysis by using the beta plane approximation \cite{shukla2003zonal}. 
The equilibrium velocity profile of an two dimensional incompressible
shear flow is represented by hyperbolic-tangent profile that is subjected to external differential rotation. The effects of the linear variation in the rotational frequency along the perpendicular to the flow direction are incorporated by using the beta plane approximation.   
%
%
 %
The analytical treatment of instability in a horizontally sheared flow subjected to external rotation has been extensively discussed in the literature \cite{kuo1949dynamic,rayleigh1880stability,read2020baroclinic}. Stability conditions for a sheared flow in the beta plane approximation were derived using the linearized vorticity equation for rotating fluids, known as the Rayleigh-Kuo equation\cite{kuo1949dynamic}. A necessary condition for instability was established, generalizing the classical Rayleigh criterion for non-rotating fluids\cite{rayleigh1880stability}. The qualitative validation of these conditions using the pseudo-spectral numerical simulation has been presented in some detail. \\  
Studies on the physical effects of magnetic fields on KHI instability suggest that the magnetic field reduces the growth rate of the most unstable mode and tends to produce compressed and elongated vortex structures \cite{liu2018physical,keppens1999growth}. Their interpretation is given in terms of the Lorentz force, which provides magnetic tension and pressure to the vortex flow. However, it has been observed that the long-wavelength modes are not suppressed even with a very high magnetic field. In contrast to a uniform magnetic field, a non-uniform magnetic field increases the growth rate of KHI compared to the pure hydrodynamic case \cite{keppens1999growth}.
In magnetized sheared flow scenarios, the Alfven Mach number was found to be a crucial parameter in determining the stability conditions for Kelvin-Helmholtz flow \cite{liu2018physical,keppens1999growth}. In the case of rotation, the parameter $\beta$ determines the impact of rotation on the growth rate of the KH instability. Both parameters have shown a tendency to suppress the growth rate of KH modes, but they act on opposite scales. The Alfven Mach number affects shorter wavelength perturbations, while the Coriolis parameter $\beta$ acts on longer wavelength perturbations. Similar effects of compression and elongation have been observed in the presence of rotation, where the vortex is compressed and elongated due to the Coriolis Force. \\
The article is organized as follows: The model equations for a rotating incompressible flow are discussed in Sec.\ref{Model}, along with the simulation details. The linear  dispersion relation of Rossby wave is presented in Sec.\ref{Linear dispersion}. Sec.\ref{KHI_rotation} presents the Kelvin Helmholtz (KH) instability in both rotating and non-rotating sheared flows. To check the reliability of the code, the KHI results in a non-rotating frame $(\Omega = 0)$ are presented in Sec.\ref{special_case}. Sec.\ref{KHI_Finite_rotation} presents the KHI results with finite rotation  $(\Omega \neq 0)$ using both the analytical approach in Sec.\ref{analytical} and the numerical approach in Sec.\ref{General_KHI}. The analogy between rotation and magnetic field effects on KHI is established in Sec.\ref{Mag_rotation}. Conclusions drawn from the analysis are presented in Sec.~\ref{conclusion}.

\section{Model Equations for a rotating incompressible flow}\label{Model}
%
The paper aims provide significant insights into the underlying physics governing the Kelvin-Helmholtz instability of a rotating fluid. The KH instability is frequently observed at the boundary where two adjacent fluid streams exhibit velocity gradients. The dynamics of an incompressible two-dimensional fluid flow subjected to rotation are governed by the Euler equations in a rotating frame, which are given as,  
 \begin{equation}
  \nabla\cdot\textbf{u}_d = 0,\label{continuity}
 \end{equation} 
 and %
%
 \begin{equation}
 \frac{\partial\textbf{u} }{\partial t} + (\textbf{u}\cdot\nabla) \textbf{u} = 2\textbf{u}\times\boldsymbol{\Omega} - \nabla P+ \nu \nabla^{2} \textbf{u}. \label{momentum}
\end{equation}
The Eq.~\ref{continuity} and \ref{momentum} are the mass and momentum conservation equations of a rotating fluid, respectively. The $\textbf{u}$, $P$, and $\nu$ are the flow velocity, pressure, and viscosity of the fluid. The symbol $\Omega = \Omega_{0} \uvec{z}$ represents the rotational frequency, with $\Omega_{0}$ denoting its magnitude and $\uvec{z}$ indicating the direction perpendicular to the plane containing the fluid flow.   It is convenient to express these equations in the vorticity-streamfunction formulation where the velocity of the fluid is described in terms of potential flows. The above equations in vorticity-streamfunction
formalism become, 
 \begin{equation}
\frac{\partial \boldsymbol{\omega_{z}}}{\partial t} + (\textbf{u}\cdot\nabla) \boldsymbol{\omega_{z}} = \nabla \times \left(2\textbf{u}\times\boldsymbol{\Omega}\right) - \nu \nabla^{2} \boldsymbol{\omega_{z}}.  \label{Vorticity}
\end{equation}
 \begin{equation}
\nabla^{2} \psi = - \boldsymbol{\omega_{z}}. \label{Possion_eq}
\end{equation}   
The flow velocity component can be determined as $u_{x} = \partial \psi/ \partial y$ and  $u_{y} = -\partial \psi/ \partial x$. 
Using the standard vector identity for the curl of a vector cross product, we write,
\begin{equation}\label{vector_idenity}
\begin{split}
\nabla \times (\textbf{u} \times \boldsymbol{\Omega}) = \textbf u(\nabla \cdot \boldsymbol{\Omega})-\boldsymbol{\Omega}(\nabla\cdot\textbf{u}) + \boldsymbol{\Omega}\cdot\nabla\textbf{u}\\ -(\textbf{u}\cdot\nabla)\boldsymbol{\Omega}.
\end{split}
\end{equation} 
We note that the first, second and third term of right hand side of 
Eq.~\eqref{vector_idenity} either vanish or negligible for our setup.  
Specifically, the first term vanishes because rotational frequency is 
constant along $z$-direction, $\nabla\cdot \boldsymbol{\Omega} = 0$, 
while the second term approaches zero because fluid flow 
is assumed to incompressible ($\nabla\cdot\textbf{u}_d = 0$). 
The third term vanishes because there is no gradient in $\textbf{u}$ along 
$\boldsymbol{\Omega}$ either.  
Under these conditions and $\Omega$ = $\Omega_{0}$ $\uvec{z}$ , Eq.~\eqref{vector_idenity} reduces to,
\begin{equation}\label{approximation}
\begin{split}
\nabla \times (\textbf{u} \times \boldsymbol{2\Omega}) = -(\textbf{u}\cdot\nabla){2\Omega} \uvec{z}  .
\end{split}
\end{equation}
We consider the two-dimensional rotating fluid with
2$\Omega$ = (2$\Omega_{0}$  $\sin$ $\lambda$) $\uvec{z}$ =   $f$ $\uvec{z}$ , where $\lambda$ is the latitude of the site counted from the equator on the
planet, the $x$, $y$, and $z$ axes are directed eastwards, northwards, and upwards, respectively, $f$ is
the Coriolis frequency, $f$ = $f_{0}$ + $\beta$y, where $f_{0}$ =  2$\Omega_{0}$ $\sin$ $\lambda_{0}$ and $\beta$ = (2$\Omega_{0}$/$r_{0}$) $\cos$ $\lambda_{0}$ $>$ 0.
Here, within the $\beta$-plane approximation, $\lambda_{0}$ is the latitude of the site, $r_{0}$ is the distance from the
center of the planet, and $y$ = $r_{0}$($\lambda$ - $\lambda_{0}$)\cite{shukla2003zonal}. 
\subsection*{Simulations Details}\label{Simulation}
%
%
%
Under the beta-plane approximations, the Eq.~\ref{approximation} reduces to a simple form which is given as,
\begin{equation}\label{approximation1}
\begin{split}
\nabla \times (\textbf{u} \times \boldsymbol{2\Omega}) = \beta  \frac{\partial {\psi}}{\partial x},
\end{split}
\end{equation}
Substituting this into Eq.~\ref{Vorticity}, we obtain the vorticity equation written in a rotating frame under the beta plane approximation, which is given as,
\begin{eqnarray}
 \frac{\partial \vec{\omega_{z}}}{\partial t} + J[\psi, \omega_{z}] - \beta \frac{\partial {\psi}}{\partial x} = \nu \nabla^{2} \omega_{z}\label{eq:vorticity_beta} 
 \end{eqnarray}
%
%
In the following analysis, we employ typical large-scale setup values for length, velocity, frequency, and other relevant quantities to scale the parameters and variables involved in the formulation.
Specifically, we use the values of radius of a  planet $r_{0}$, acoustic frequency ($\omega_{A}$), and the acoustic velocity $U_{a}$ typically found in a large scale setup to normalize our variables. This normalization ensures that the variables $\omega_{z}$ and $\psi$ have the unit $U_{a}$/$r_{0}$ and $U_{a}r_{0}$, respectively.
The parameters viscosity $\nu$ is also normalized accordingly. It is important to note that the Jacobian is defined as $J[\psi, \omega] = \partial_{x}\psi \partial_{y}\omega - \partial_{x}\omega \partial_{y}\psi$. \\ \\
To simulate the dynamics of the rotating fluid governed by Eqs.~\ref{eq:vorticity_beta} and \ref{Possion_eq}, we employ a pseudo-spectral technique using the Fast Fourier Transform (FFTW) library \cite{FFTWgen99}. The simulation has a resolution of $N_{x}\times N_{y}$ = 512 $\times$ 512 and a time step interval of $\delta$t = $10^{-3}$. In our numerical approach, we use variables 
$k$ for spatial discretization and $t$ for temporal discretization. These discretization techniques are chosen to satisfy the Courant-Friedrichs-Lewy (CFL) condition\cite{russell1989stability}. For time-stepping, we implement the Adams-Bashforth method \cite{durran1991third}.  The nonlinear term, for example, the 2nd term on the left-hand side of Eq.~\ref{eq:vorticity_beta} give rise to aliasing effects, which are removed by using the most accessible two-thirds rule introduced by Orszag\cite{patterson1971spectral}.
\section{Linear Analysis and Rossby dispersion relation}\label{Linear dispersion}
A low-amplitude perturbation in $\omega_{z}$ and $\psi$ is first considered to obtain a linear dispersion relation of the Rossby wave with different values of the Rossby parameter $\beta$. The perturbations in $\omega_{z}$ and $\psi$ are equal, with values $\omega_{z}/(U_{A}/r_{0}) = 0.001$.
In the linear approximation, we can neglect the nonlinear Poisson bracket terms in Eq.~\ref{eq:vorticity_beta} and Fourier transform the resultant equations by considering the stream functions to be proportional to $\exp(-i\omega t + i k \cdot r)$, where $\omega$ and $k$ are the frequency and the wave-vector, respectively. Combining the Fourier-transformed equations, the linear dispersion relation can be obtained which is given as,
\begin{figure}[ht]\centering
\includegraphics[width=\linewidth]{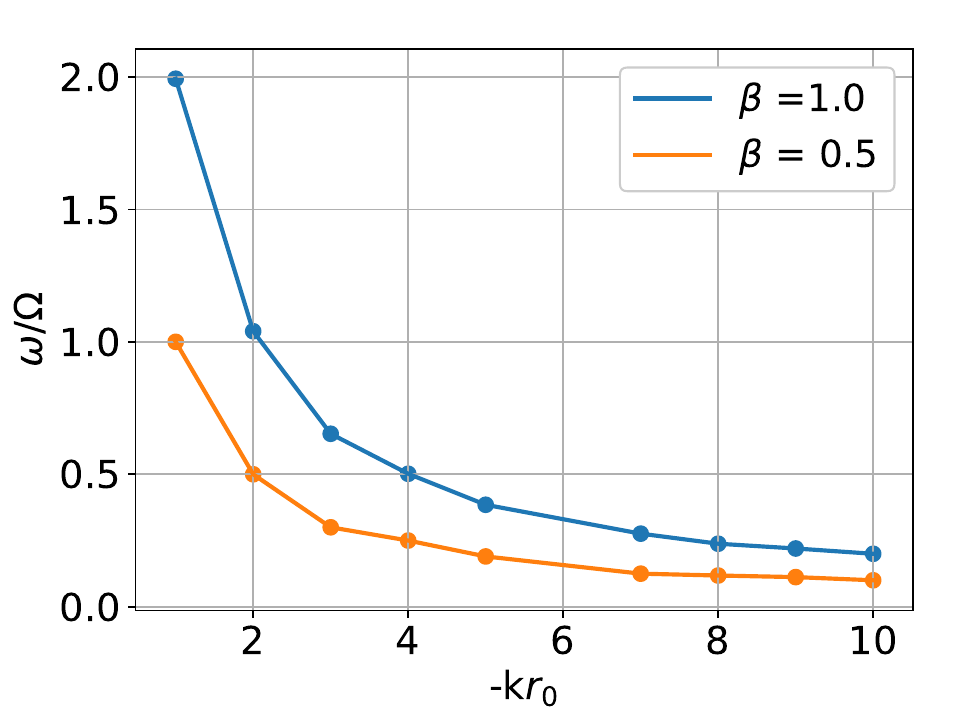}
\caption{The figure represents the linear dispersion relation of the Rossby wave for different values of $\beta$. The  yellow and blue points are calculated for $\beta$ = 0.5 and 1.0, respectively.}
\label{fig:Rossby_wave}
\end{figure}
%
%
\begin{equation}
\omega^2 + \omega \left( \frac{\beta k_{x}}{k^{2}_{x}+k^{2}_{y}}\right) = 0,
\end{equation}
%
%
The no-trivial solution for $k_{y} = 0$ of the equation is given as, 
\begin{equation}
\omega = -\left( \frac{\beta}{k_{x}}\right).\label{wave_dispersion}
\end{equation}
The wave dispersion relation presented Fig.~\ref{fig:Rossby_wave} for different value of $\beta$ parameter.  The yellow and blue dots represent the normal modes of the Rossby wave for $\beta = 0.5$ and $1.0$, respectively. It is shown in Fig.~\ref{fig:Rossby_wave} that the frequency of the wave increases with increasing the value of $\beta$. It is clear from Eq.~\ref{wave_dispersion} that the Rossby wave travels in the negative $x$-direction (east to west) at a faster rate for long-wavelength excitations and slows down as the wave vector increases, eventually reaching an almost null saturated frequency. These linear characteristic features of the Rossby wave would be useful for describing the nonlinear phenomena in a rotating fluid.
%
%
%
\section{Stability analysis of a Sheared Flow subjected to differential rotation}\label{KHI_rotation} 
\subsection{KH results with zero rotation ($\Omega$ = 0)}\label{special_case}
Sec.\ref{special_case} is devoted to obtaining the Kelvin-Helmholtz results in the absence of rotation ($\Omega = 0$). Under this assumption, Eq.\ref{eq:vorticity_beta} reduces to a simple form describing the dynamics of a non-rotating, incompressible fluid flow. The set of fluid equations is given as,
\begin{eqnarray}
 \frac{\partial \vec{\omega_{z}}}{\partial t} + J[\phi, \omega_{z}] = \nu \nabla^{2} \omega_{z}\label{eq:incom_vorticity} \\
 \nabla^{2}\psi = -\omega_{z}, \label{eq:Possion} 
 \end{eqnarray}
where $\psi$ represents the streamfunction suitably describes the dynamics of a two-dimensional incompressible fluid flow.
\begin{figure}[ht]\centering
\includegraphics[width=\linewidth]{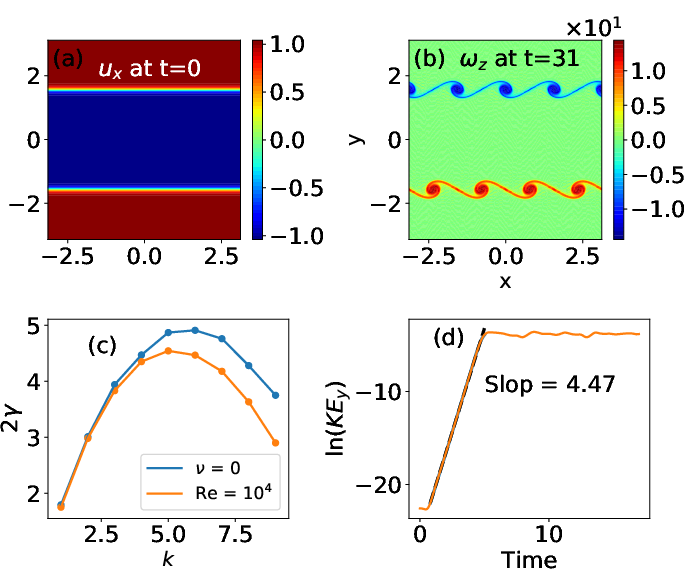}
\caption{ Figures (a) and (b) represent the initial equilibrium non-rotating sheared flow velocity perturbed with mode $k_{x} = 4$ and a snapshot of its time evolution at $t=5 t_{N}$, respectively. Figures (c) and (d) represent the growth rate of different wavelengths or modes for the different Reynolds numbers and time evolution logarithmic of the kinetic energy along $y$-direction, respectively.}
\label{KHI_HD}
\end{figure}
We consider a 2D periodic system with smooth initial vorticity profile given as, 
 \begin{eqnarray}
\omega_{z} = \frac{\omega_{0}}{\cosh^2\left(\frac{y+L_{y}/4}{d}\right)}- \frac{\omega_{0}}{\cosh^2\left(\frac{y-L_{y}/4}{d}\right)}\nonumber \\ + \frac{\omega_{0}}{\cosh^2\left(\frac{y-3L_{y}/4}{d}\right)}- \frac{\omega_{0}}{\cosh^2\left(\frac{y+3L_{y}/4}{d}\right)}\label{Initial_Profile},
 \end{eqnarray}
where the $L_{y}$, $\omega_{0}$ and $d$ are the system length, intensity of the vorticity and sheared flow width, respectively. The corresponding initial velocity $u_{x}$ profile is plotted in Fig~\ref{KHI_HD}(a) by using constant values of $\omega_{0}$ = 13 and wave-vector $k_{x}$ = 4.

The simulations are performed using an in-house developed pseudo-spectral code with parameters $N_{x} = N_{y} = 512$, $L_{x} = L_{y} = 2\pi$, and $d = 0.078$. Figs.\ref{KHI_HD}(a), (b), and (d) represent the x-component of velocity ($u_{x}$), vorticity ($\omega_{z}$) profile at $t = 31.4 t_{N}$, and natural log of the $y$-component of energy ($E_{y}$) vs. time, respectively. In order to analyze the growth rate of KHI, we impose a perturbation to the vorticity at $t = 0$ of the form $\omega_{z} = A \cos(k_{x}x)$, where $A = 0.001 \ll \omega_{0}$ is the initial amplitude of the perturbation, and $k_{x} = 4$ is its wave number along the $x$-direction. Fig.\ref{KHI_HD}(c) represents the growth rate of the individual perturbation and shows agreement with the previously known results on the KHI. These results confirm the reliability of our code for exploring the KHI in a more complex geometry. Fig.~\ref{KHI_HD} represents the growth rate of different wavelengths or modes for two values of the viscosity. The blue and yellow dots are calculated for $\nu = 0$ and $\nu \neq 0$, respectively. It can be concluded from Fig.~\ref{KHI_HD} that viscosity affects only the long-wavelength modes. 
\subsection{KH results with finite rotation ($\Omega \neq 0$)} \label{KHI_Finite_rotation}
\subsubsection{Analytical treatment of Perturbations on a Sheared flow in a rotating frame} \label{analytical}
In Sec.~\ref{analytical}, we consider an analytical approach to determine the stability conditions of a step-wise shear flow in a rotating frame. 
Starting with the vorticity equation\cite{howard1964instability,engevik2004note,burns2002barotropic}:
\begin{equation}
\frac{\partial \omega}{\partial t} + (\mathbf{U} \cdot \nabla) \omega = \left(\beta - \frac{d^2 \mathbf{U}}{dy^2}\right) \frac{\partial \psi}{\partial x} + \nu \nabla^2 \omega \label{Beta_Plane} 
\end{equation}
After using the relation $\nabla^{2}\psi = -\omega$, the Eq.~\ref{Beta_Plane}
can be re-write in term of streamfunction which is given as,
\begin{equation}
\frac{\partial (\nabla^2 \psi)}{\partial t} + (\mathbf{U} \cdot \nabla) \nabla^2 \psi =\left(\beta - \frac{d^2 \mathbf{U}}{dy^2}\right)\frac{\partial \psi}{\partial x} + \nu \nabla^2 (\nabla^2 \psi). \label{Beta_Streamfunction} 
\end{equation}
It is mathematically convenient to ignore the viscous effects from the Eq.~\ref{Beta_Streamfunction} such that it reduces to simple form which is given as, 
\begin{equation}
\frac{\partial (\nabla^2 \psi)}{\partial t} + (\mathbf{U} \cdot \nabla) \nabla^2 \psi -\left(\beta - \frac{d^2 \mathbf{U}}{dy^2}\right) \frac{\partial \psi}{\partial x} =0, \label{Beta_no_viscosity} 
\end{equation}
for a initial velocity profile $U = U(y)\uvec{x}$.
As the objective of this linear analysis is to identify the instability criteria under which the fundamental flow subjected to rotation or beta plane effect, the focus is solely on two-dimensional perturbations.
Using normal mode analysis in which physical quantities, for example, $\psi$ can be expressed as,
 \begin{eqnarray}
\psi(x,y,t) = \tilde{\psi}(y) \exp[(i(kx-\omega t)], \label{Normal_mode} 
\end{eqnarray}
where $k$ and $\omega$ are the $k_{th}$ mode and corresponding frequency of a perturbation at the interface. Introducing this normal mode ansatz into
Eq.~\ref{Normal_mode} yields, 
 \begin{eqnarray}
\frac{\partial^2 \tilde{\psi}(y)}{\partial y^2}-\left(k^2-\frac{\beta-\frac{d^2 \mathbf{U}}{dy^2}}{(U-c)}\right)\tilde{\psi}(y)= 0, \label{differential_equation} 
\end{eqnarray}
associated to vanishing boundary conditions. The Eq.~\ref{differential_equation} has a solution of the form given as,
 \begin{eqnarray}
\psi_{j} = A_{j} \exp^{(-\kappa y)} + B_{j} \exp^{(\kappa y)}, \label{solution} 
\end{eqnarray}
where $\kappa$ = $\sqrt{\left(k^2-\frac{\beta-\frac{d^2 \mathbf{U}}{dy^2}}{(U-c)}\right)}$. The presence of a imaginary part in the phase velocity $c$ guarantees the existence of a growing disturbance  and thus the instability of the basic flow. Conversely, the basic flow is stable if and only if the phase speed is purely real. Because it is impossible in general to determine the values $c$ for an arbitrary velocity profile, it was attempted to solve to reach a weaker stability criterion. We investigate streamwise wavenumbers $k$ that fall within the positive real number domain. Demonstrably, as $y$ tends towards negative (or positive) infinity, the general solution (given by Eq.~\ref{solution}) experiences divergence. 
These are the nonlinear characteristic-value problem with parameter $c$. The general solutions of the equation had been discussed by Kuo et al. in Ref.~\cite{kuo1949dynamic}.  The instability of the flow was predicted to occur only when the term $\beta - d^2 U/dy^2$ either vanishes or changes sign within the domain \cite{engevik2004note,burns2002barotropic}. 
It has been observed that the phase velocity $c_{r}$ is negative for all $k$ and $\beta$, indicating that the waves propagate toward the west. The absolute value of $c_{r}$ increases with $\beta$ for a fixed $k$ and decreases with $k$ for a fixed $\beta$. It has also been observed that for $\beta = 0$, the maximum $c_{i}$ occurs at $k = 0$, but it shifts toward a larger $k$ as $\beta$ increases. For a fixed $k$, $c_{i}$ decreases with increasing $\beta$, indicating the stabilizing influence of $\beta$ \cite{engevik2004note,burns2002barotropic}.

More specifically, the stability conditions of normal mode perturbations in a parallel shear flow subjected to rotation under the beta plane approximation are determined by $\beta$ and the curvature in the flow velocity profile. In the absence of rotation ($\beta = 0$) or when $\beta = d^2 U/dy^2$, the above equations have solutions indicating that all modes are unstable for finite shear in the flow velocity ($\delta U \neq 0$). The inclusion of $\beta$ in the equations results in coupled equations, where $\kappa$ quantifies the impact of $\beta$ on the stability conditions of normal modes. The significance of the $\beta$ term on shorter perturbation modes can be determined from the expression of $\kappa$, particularly when the term $\left(\frac{\beta}{(U-c)}\right)$ becomes comparable to the perturbation mode number $k$. For cases where $\beta > d^2 U/dy^2$, the term $\beta-d^2 U/dy^2$ remains positive throughout the flow domain, indicating stability of the equilibrium flow against all mode perturbations. 


\subsubsection{Stability analysis with a more general velocity profile: A numerical approach} \label{General_KHI}
The qualitative description of the stability characteristics of the shear flow has revealed that rotation diminishes the instability of the flow and acts only on longer wavelength perturbations. Sec.~\ref{General_KHI} focuses on investigating the stability characteristics of a vorticity profile given in Eq.~\ref{Initial_Profile}, which offers a more realistic representation of the flow. 
Due to the challenges of the analytical analysis, we employ numerical simulations to investigate the effect of rotation on the stability of a velocity profile given by $U = U_{0} \tanh(y/d)$, where $U_{0}$ and $d$ represent the initial velocity magnitude and the smoothing length, respectively.  \\
\begin{figure}[ht]\centering
\includegraphics[width=\linewidth]{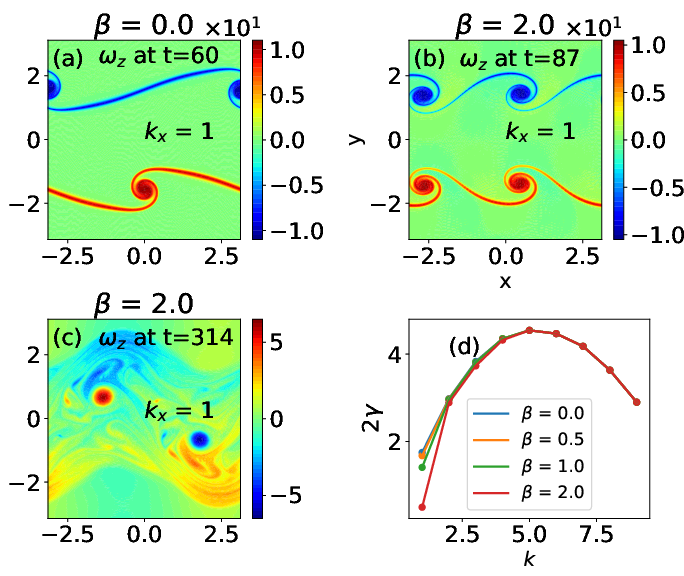}
\caption{Figure (a) represents a snapshot of the time evolution of the imposed mode $k_{x} = 1$ on the initial vorticity profile ($W_{0} = 12$) at $t=69.08 t_{N}$ for $\beta = 0$. Figures (b) and (c) represent snapshots of its time evolution with $\beta = 2.0$ at different times, $t=87.92 t_{N}$ and $t=314 t_{N}$, respectively. Figure (d) represents the growth rate of different wavelengths or modes for various values of $\beta$.}
\label{w_13}
\end{figure}
The simulations are performed using the following parameters: $N_{x} = N_{y} = 512$, $L_{x} = L_{y} = 2\pi$, $\omega_{0} = 13.0$, $U_{0} = 2.0$ $d = 0.078$, mode $m = 1$, and $R = 10^{4}$. Figs.~\ref{w_13} (a), (b), and (c) represent the vorticity ($\omega_{z}$) profile at $t = 69.08 t_{N}$ with $\beta = 0.0$, at $t = 87.92 t_{N}$ with $\beta = 2.0$, and at $t = 314 t_{N}$ with $\beta = 2.0$, respectively. The formation of two vortex structures, as seen in Fig.~\ref{w_13} (b), indicates the redistribution of energy from the initial mode $m=1$ to the second mode $m=2$. This redistribution is further confirmed by the mode number spectra of kinetic energy $E_{y}$ obtained in the presence of rotation (spectra not shown in this figure). Initially, energy cascades from  $m=1$ to $m=2$, and then it grows over time before ultimately merging into a single vortex structure, as depicted in Fig.~\ref{w_13}(c). The growth rate profile of normal modes is plotted in Fig.~\ref{w_13}(d) for four different values of the $\beta$ parameter. The growth rate of all modes remains unchanged for all values of $\beta$, except for mode $m=1$. This observation aligns with the analytical findings presented in the preceding section, indicating that the greater sheared flow velocity remains unaffected by rotation until a significantly high value of $\beta$ is reached. In order to demonstrate the significant influence of $\beta$ on sheared flow instability, it is necessary to examine scenarios in which the sheared flow velocity is lower and the $\beta$ parameter is higher. \\
\begin{figure}
\includegraphics[width=\linewidth]{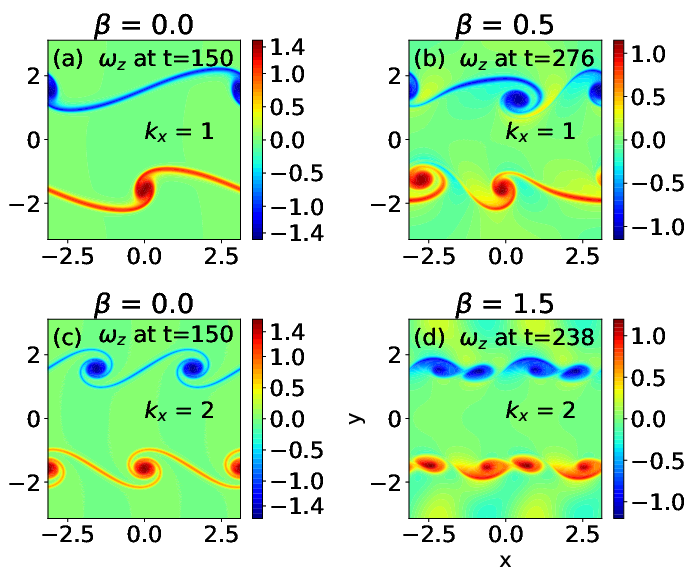}
\caption{Figures (a) and (b) represent a snapshot of the time evolution of the imposed mode $k_{x} = 1$ on the initial vorticity profile ($W_{0} = 2.0$) for $\beta = 0.0$ and 0.5, respectively. Figures (c) and (d) represent a snapshot of the time evolution of the imposed mode $k_{x} = 2$ on the initial vorticity profile ($W_{0} = 2.0$) for $\beta = 0.0$ and 1.5, respectively.}
\label{w_2}
\end{figure}
\begin{figure}
\includegraphics[width=\linewidth]{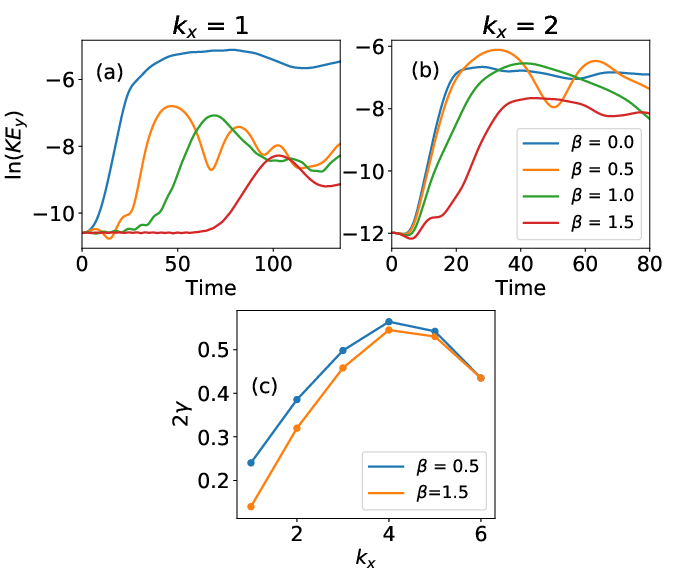}
\caption{Figures (a) and (b) represent the time evolution logarithm of the kinetic energy along the y-direction for $k_{x} = 1$ and $k_{x} = 2$, respectively, with different values of $\beta$. Figure (c) represents the growth rate of different wavelengths or modes for different values of $\beta$.}
\label{growth}
\end{figure}
Therefore, we present the characteristics of KH vortices in the presence of relatively lower shear flow velocity and  higher $\beta$ values. Figs.~\ref{w_2} (a) and (b) represent vorticity ($\omega_{z}$) profile at $t = 150.72 t_{N}$ with $\beta = 0.0$, and at $t = 276.32 t_{N}$ with $\beta = 0.5$, respectively. The perturbation mode $m=1$ and the initial flow velocity $U_{0}$ = $0.15$ are considered for generating these plots.
Again, the formation of two KH vortex structures, as seen from Fig.~\ref{w_2} (b), indicates that the energy has been redistributed from
the initial mode $m=1$ to second mode $m=2$. Figs.~\ref{w_2}(c) and (d) represents the vorticity ($\omega_{z}$) profile at $t = 150.72 t_{N}$ with $\beta = 0.0$ and $m=1$, at $t = 238.64 t_{N}$ with $\beta = 0.5$ and $m=2$, respectively. Now, the formation of four KH vortex structures, as seen from Fig.~\ref{w_2} (d), indicates that the energy has been redistributed from
the initial mode $m=2$ to the fourth mode $m=4$. It is important to observe that the four vorticity patches that form exhibit anisotropic characteristics. We present a qualitative description of these observations in the upcoming section. 

The growth rate profiles plotted in Fig.~\ref{growth} show that the growth rates of the individual modes are significantly altered even at the lower value of $\beta = 0.5$. This observation again aligns with the analytical findings presented in the preceding section, indicating that the lower sheared flow velocity is highly affected by rotation. 
The growth rate plot of $E_{y}$ in Fig.~\ref{growth} illustrates that the perturbed velocity undergoes exponential growth as time advances, eventually reaching saturation in the nonlinear regime. The power associated to the perturbed velocity field corresponds to the average energy of the system. The tracking of this quantity as a function of time provides a good measure of the growth of the instability and its saturation in the nonlinear regime.
We show the evolution of $E_{y}$ as
a function of time in the semilog plot in Fig.~\ref{growth} (a) and (b) for the modes $m =1$ and $m=2$, respectively. The figures clearly demonstrate that as $\beta$ increases, the growth of the perturbed mode diminishes. In order to impact higher modes, a larger $\beta$ value becomes essential for a fixed value of equilibrium flow velocity. These observations qualitatively agree with the theoretical predictions presented in the previous section. In the upcoming section, we explore the physical effects of rotation on the Kelvin-Helmholtz instability and draw analogies with the effects of magnetic fields.
%
\section{Characterizing  KH flows under Rotation: A Qualitative Comparison with Magnetic Fields }
\label{Mag_rotation}
\begin{figure}[ht]\centering
\includegraphics[width=\linewidth]{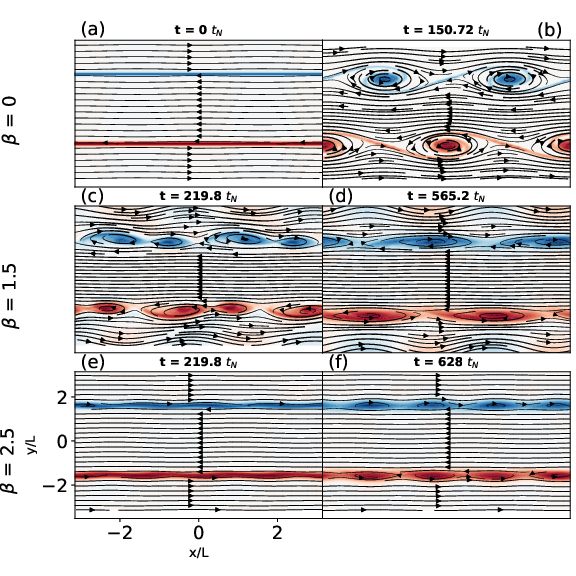}
\caption{Snapshots of the streamline and vorticity of the sheared flow initially perturbed with mode $k_{x} = 2$ at different times. Each row corresponds to a different value of the parameter $\beta$. Figures (a) and (b) are for $\beta=0$, (c) and (d) are for $\beta=1.5$, and (e) and (f) are for $\beta=2.5$.}
\label{k2}
\end{figure}

It has been already predicted that the magnetic field parallel to flow direction contributes in suppressing the instability of the flow against the small perturbation\cite{liu2018physical,keppens1999growth}. The stabilizing effect of the magnetic field is known to originate from the Lorentz force. As the Coriolis force mimics the effects of the Lorentz force in a rotating frame, the purpose of Sec.~\ref{Mag_rotation} is to understand the role of the Coriolis force in situations analogous to magnetic field cases. The force exerted by the rotation ($\Omega \uvec{z}$) in the beta-plane approximation is directed along the eastward direction or $x$-direction, as the variation of the Coriolis parameter is assumed to be along the $y$-direction or meridional direction. The Coriolis effect directs the flow in the negative $x$-direction, given that $\beta$ consistently remains positive.
In magnetized sheared flow scenarios, the Alfven Mach number was found to be a crucial parameter in determining the stability conditions for  Kelvin-Helmholtz flows\cite{liu2018physical,keppens1999growth}. In the case of rotation, the parameter $\beta$ determines the impact of rotation on the growth rate of the KH instability. Both parameters have shown a tendency to suppress the growth rate of KH modes, but they act on opposite scales. The Alfven Mach number affects shorter wavelength perturbations, while the Coriolis parameter $\beta$ acts on longer wavelength perturbations.
%
%
\begin{figure}[ht]\centering
\includegraphics[width=\linewidth]{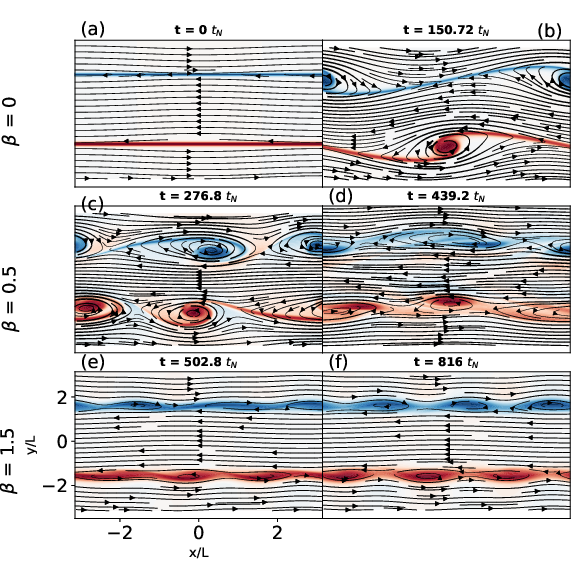}
\caption{Snapshots of the streamline and vorticity of the sheared flow initially perturbed with mode $k_{x} = 1$ at different times. Each row corresponds to a different value of the parameter $\beta$. Figures (a) and (b) are for $\beta=0$, (c) and (d) are for $\beta=0.5$, and (e) and (f) are for $\beta=1.5$.}
\label{k1}
\end{figure}
The linear growth of the $y$-component of the energy, as depicted in Figure~\ref{growth}(a) and (b), illustrates that the growth rate decreases with increasing $\beta$ parameter. The effects of the $\beta$ parameter, as plotted in Fig.~\ref{growth}(c), are noticeable at smaller values of $k$, while higher modes are insensitive to the $\beta$ parameter. In magnetized Kelvin-Helmholtz flow, the narrow spectral space of unstable modes ($0<k<1$) persists even in very high magnetic fields\cite{liu2018physical}. However, this region of unstable modes disappears even with a small value of the $\beta$ parameter.  
%
To visualize the effect of rotation on the KHI, the streamlines along with vorticity at different times under rotation of different strengths are presented in Fig.~\ref{k1} and Fig.~\ref{k2}, for mode number $k = 1$ and $k=2$, respectively. The initial equilibrium sheared flow, subjected to an imposed perturbation with $k_{x} = 1$, is depicted in Fig.\ref{k1}(a). Subsequently, it rolls up into a billow, grows, and eventually forms a single vortex in the absence of rotation, indicating thorough fluid mixing, as illustrated in Fig.\ref{k1}(b). As shown in Fig.~\ref{k1}(c) and (d), for $\beta = 0.5$, two vortices emerge, which are compressed and elongated along the zonal direction as a result of the combined effect of rotation and mean flow. Similar effects have been observed in a magnetized sheared flow, where the vortex is compressed and elongated due to the combination of magnetic field and mean flow. However, the formation of two vortices has not been observed in the presence of a magnetic field.  In Fig.~\ref{k1}(e) and (f), the vortices appear to be more compressed and elongated for $\beta$ = 1.5, suggesting that the destabilizing influence starts to diminish in the presence of rotation. The emergence of the vortex sheet, characterized by compressed flat vortices confined in shear flow, occurs at a $\beta$ value of 2.5 (not depicted here). These results resemble magnetized sheared flow, where a similar vortex sheet emerged at an Alfven Mach number of 2.15\cite{liu2018physical}.

%
The Fig.~\ref{k1} and Fig.~\ref{k2} presents the vorticity field, which provides valuable kinetic information about the flow. The red and blue colors represent positive and negative vorticity, respectively. 
From Fig.~\ref{k1}(c), it is evident that the region between the pair of positive and negative vortices is associated with opposite vorticity, which diminishes over time. The distribution of negative vorticity wrapped around the outside of the elongated vortex is more pronounced for the magnetized flow\cite{liu2018physical}. This is because magnetic field lines are folded in different directions by fluid vorticity. \\
%
%
The physical explanation of how the Coriolis parameter $\beta$ affects the Kelvin-Helmholtz instability (KHI) can be described by the sole involved term, expressed as $\beta u_{y}$. The term represents the external rotation that varies linearly along the $y$-direction, with its gradient defined by the parameter $\beta$. A flow in the $y$-direction encounters the influence of the Coriolis force that acts perpendicular to the flow, and its magnitude is determined by the $\beta$ parameter.  However, the $x$-direction or zonal flow does not experience the effect of Coriolis force, as the Coriolis parameter is assumed to vary along the $y$-direction only. An interesting observation from Fig.~\ref{k1} is that the flow along the negative $x$-direction appears to be largely insensitive to the Coriolis force, resulting in the adjacent vortex flow becoming flattened. Conversely, flows directed along the positive $x$-direction allow the vortex flow to exhibit curvature. This effect is visible from Fig.~\ref{k2} plotted for showing the evolution of sheared flow perturbed with $m=2$ with different values of $\beta$. 
As shown from Fig.~\ref{k2}(c) to (f), four vortices emerge, which are compressed and elongated along the zonal direction as a result of the combined effect of rotation and mean flow. However, to suppress the instability, a relatively higher value of $\beta$ is required compared to the mode $(m) = 1$ cases.
\section{Conclusions}\label{conclusion}
We have performed a pseudo-spectral simulation to study the effect of rotation on Kelvin-Helmholtz flows at high Reynolds numbers. The equilibrium velocity
profile of an two dimensional incompressible shear flow is
represented by a hyperbolic-tangent profile. The rotation is assumed to be perpendicular to the plane containing the equilibrium flow and to vary across the flow. The variation of the rotational frequency is approximated using the beta-plane approximation. The parameter $\beta$, defining the strength of the Coriolis force, exhibits finite effects on both the growth rate of Kelvin-Helmholtz instability and the formation of nonlinear vortices, analogous to the Alfven parameter in magnetized shear flows. The numerical results show a close qualitative agreement with the analytical results obtained for a step-wise shear flow profile. The analytical results suggest that the higher sheared flow velocity remains unaffected by rotation until $\beta$ reaches a significantly high value. It is also predicted that the impact of $\beta$ is significant on longer-wavelength perturbations, whereas shorter wavelengths remain unaffected by $\beta$.

The magnetic field and rotation have shown a tendency to reduce the growth rate of the KHI. However, the magnetic field and Coriolis force act on different scales, with the latter suppressing
long-wavelength mode perturbations. The long-wave perturbations that need a strong magnetic field are readily suppressed by rotation. 
Similar to the magnetic field case, the Coriolis force suppresses KHI, and tends to form compressed and elongated KH
vortex structures. A higher number of vortices are observed in the presence of rotation compared to non-rotating cases. The rotation has shown a tendency to redistribute energy among the modes. The energy from mode $m=1$ is transferred to even-numbered modes like $m=2$, and from $m=2$ to $m=4$.  This results in the formation of two and four vortex structures when perturbations are given in $m=1$ and $m=2$, respectively. However, the formation of multiple vortices has not been observed in the presence of a magnetic field.  
The physical explanation of how the Coriolis parameter $\beta$ affects the Kelvin-Helmholtz instability (KHI) can be described
by the sole involved term, expressed as $\beta$ $u_{y}$. The term represents the external rotation that varies linearly along the $y$-
direction, with its gradient defined by the $\beta$ parameter. A flow
in the $y$-direction encounters the influence of the Coriolis force
that acts perpendicular to the flow, and its magnitude is determined by the $\beta$ parameter. However, the $x$-direction or zonal
flow does not experience the effect of Coriolis force, as the
Coriolis parameter is assumed to vary only along the $y$-direction. This paper did not consider the effect of a magnetic field. 
%
An interesting problem can be explored where both pseudo-magnetic effects (introduced by rotation) and real magnetic field effects are involved

 \bibliographystyle{apsrev4-1}
 \bibliography{paper}


\end{document}